\documentclass[a4paper,10pt,twoside]{cpc-hepnp}

\usepackage{multicol}
\usepackage{multirow}
\usepackage{graphicx}
\usepackage{booktabs}
\usepackage{amssymb,bm,mathrsfs,bbm,amscd}
\usepackage[tbtags]{amsmath}
\usepackage{lastpage}

\begin{document}

\fancyhead[c]{\small Chinese Physics C~~~Vol. 37, No. 1 (2015)
010201} \fancyfoot[C]{\small 010201-\thepage}

\footnotetext[0]{Received 15 April 2015}

\title{Structural analysis of superconducting dipole prototype for HIAF}

\author{%
      ZHANG Xiao-Ying(ÕÅÏþÓ¥)$^{1,1)}$\email{zxydream@impcas.ac.cn}%
\quad WU Bei-Min(Îâ±±Ãñ)$^{1}$
\quad NI Dong-Sheng(Ä߶«Éý)$^{1}$\\
\quad CHEN Yu-Quan(³ÂÓñȪ)$^{1;2}$
\quad WU Wei(ÎâΡ)$^{1}$
\quad MA Li-Zhen(ÂíÁ¦ìõ)$^{1}$
}
\maketitle

\address{%
$^1$ Institute of Modern Physics,  Chinese Academy of Sciences,  Lanzhou,  730000,  China\\
$^2$ University of Chinese Academy of Sciences,  Beijing,  100049,  China\\
}

\begin{abstract}
The High Intensity Heavy-Ion Accelerator Facility is a new project in the Institute of Modern Physics. The dipole magnets of all rings are conceived as fast cycled superconducting magnet with high magnetic field and large gap, the warm iron and superconducting coil structure (superferric) is adopted. The reasonable structure design of coil and cryostat is very important for reliable operation. Based on the finite element software ANSYS, the mechanical analysis of electromagnetic stress, the thermal stress in the cooling down and the stress in the pumping are showed in detail. According to the analysis result, the supporter structure is the key problem of coil system. With reasonable support's structure design, the stress and the deformation of coil structure can be reduced effectively, which ensure the stable operation of superconducting coil system.
\end{abstract}

\begin{keyword}
mechanical analysis,  electromagnetic force,  stress-strain,  thermal stress.
\end{keyword}

\begin{pacs}
41.85.Lc, 84.71.Ba
\end{pacs}

\footnotetext[0]{\hspace*{-3mm}\raisebox{0.3ex}{$\scriptstyle\copyright$}2015
Chinese Physical Society and the Institute of High Energy Physics
of the Chinese Academy of Sciences and the Institute
of Modern Physics of the Chinese Academy of Sciences and IOP Publishing Ltd}%

\begin{multicols}{2}

\section{Introduction}

The High Intensity Heavy-Ion Accelerator Facility £¨HIAF£©is a new project in the Institute of Modern Physics. In order to get the high magnetic field in the large aperture and lower the cost of operation, the superconducting dipole magnet will be used in the synchrotron accelerator\cite{lab1}. For the fast cycled dipole magnet of boost ring, its effective length is $2.6$ m, the good field region is $\pm160$ mm$\times 110$ mm and the gap is $120$ mm. With a maximal field of $2.25$ T provided through the pole yoke and with field ramp rates of $1.125$ T/s, the warm iron and superconducting coil structure is adopted\cite{lab2}, and the liquid helium inner cooling superconducting cable is used for coil winding\cite{lab3}. For the reason of special structure of coil system, some mechanical problem must be solved in the design of superconducting dipole magnet. The stress and deformation of the coil and coil case under the electromagnetic force could cause the quench, the thermal stress and deformation of coil in the cooling down and the deformation of cryostat in the pumping could damage the structure of coil and cryostat. For the reason of safety operation, the mechanical analysis is necessary for structural design of coil and cryostat system.

The structure of coil system is similar with the superconducting prototype dipole coil of super-FRS\cite{lab4}. Because the test result of the prototype dipole coil of super-FRS shown the deformation of the long edges of coil case is about $5$ mm under the electromagnetic force, it is must be avoided in the new structure design. Based on the design and test result of prototype magnet of super-FRS, there are some improvements would be done in the new structure, especially for supporters.

\section{Structure of superconducting coil system}

The Fast cycled superconducting prototype dipole coil for HIAF project consists of coil, coil case, thermal shield, cryostat and supporters. The coil is wound with liquid helium inner cooling superconducting cable based on NbTi/Cu strands, and Cu-Ni pipe with $6$ mm inner diameter and $0.5$ mm wall thickness is used for liquid helium circulation. The diameter of the cable is $10$ mm and the total turns of coil are $12$. The coils are vacuum impregnated with the epoxy resin for ground insulation and the cross section of coil is $36$ mm$\times 48$ mm after VPI. The coil case was designed for coil fixation and to protect the winding against the magnetic force during operation. The coil case and cryostat are made of $316LN$ stainless steel for the good mechanical property. An $80$ K thermal shield which will limit the heat radiation from the cryostat is placed around the coil case. The outer surface of the thermal shield will be wrapped with $30$ layers insulation to reduce the heat fluxes for helium system. The coil case, the thermal shield and cryostat are fixed and adjusted by supporters. Therefore, the structure of supporter is complex and important for coil system design.

\section{Mechanical analysis under magnetic force at nominal current}

The superconducting prototype dipole coil is trapezoid type with upper and lower coil. When the magnetic field is excited to $2.25$ T, the operation current is $11500$ A, and the expansionary force on the coil is serious. The coil case with H type is used for coil fixation, at the same time, it also used to against the deformation of coil under the electromagnetic force. Based on the sequential coupled finite element method of ANSYS, mechanical analysis is performed to get the stress and deformation distribution on the coil and coil case\cite{lab4}. In order to guarantee the accuracy of analysis, the structure model is corresponding created in OPERA and ANSYS. The node coordinates in the model are gotten firstly, and then, the magnetic field and current density output from the magnetic field analysis result in OPERA, and load it to the structure model in ANSYS to get the mechanical analysis result.

In the structural analysis, the $1/4$ model is used to make sure the constraint symmetric applying. In condition without the additional supporters, the strength of coil case is relevant to the thickness of stainless steel. The results of stress and deformation analysis with different thickness of coil case are shown in Table~\ref{tab1}. We can see that under the electromagnetic force, the effective stress (Von Mises) is concentrate in the corner of the coil case, and the general deformation of the coil and coil case is expanding (y direction) which occurred in the center of the long edges of the coil case. With increasing thickness, the effective stress and deformation of coil case are decreased as well. In the optimal structure design, $10$ mm thickness of coil case is chosen. The maximum Von Mise stress is $237$ MPa which is occur in the corner of the coil case, and the maximal deformation in the center of the long edges is about 6 mm. The large deformation of coil and coil case is not allowed in the structure design, so the extra supporters in the center of long edges are required. With the reasonable structure design of supporters, the electromagnetic force is transferred to the yoke and to decrease the deformation of the coil and coil case.

\end{multicols}

\begin{center}
    \tabcaption{\label{tab1}The stress and deformation distribution with different thickness of coil case.}
    \footnotesize
    \begin{tabular}{ccccccc}
    \hline
      \multirow{3}*{Thickness£¨mm£©} & \multicolumn{2}{c}{Maximum effective stress£¨MPa£©} & \multicolumn{4}{c}{Maximum deformation£¨mm£©} \\
    \cline{2-7}
      & \multirow{2}*{Corner} & Center of & Overall & \multirow{2}*{X direction} & \multirow{2}*{Y direction} & \multirow{2}*{Z direction} \\
      &  & long edges & deformation &  &  &  \\
    \hline
    $8$     & $304$   & $170$   & $8.036$ & $0.358$ & $8.036$ & $0.476$ \\
    $10$    & $237$   & $140$   & $6.118$ & $0.209$ & $6.118$ & $0.347$ \\
    $12$    & $190$   & $110$   & $4.807$ & $0.24$  & $4.807$ & $0.266$ \\
    $15$    & $141$   & $70$    & $3.493$ & $0.189$ & $3.493$ & $0.191$ \\
    \hline
    \end{tabular}
\end{center}

\begin{multicols}{2}

\begin{center}
\includegraphics[width=8cm]{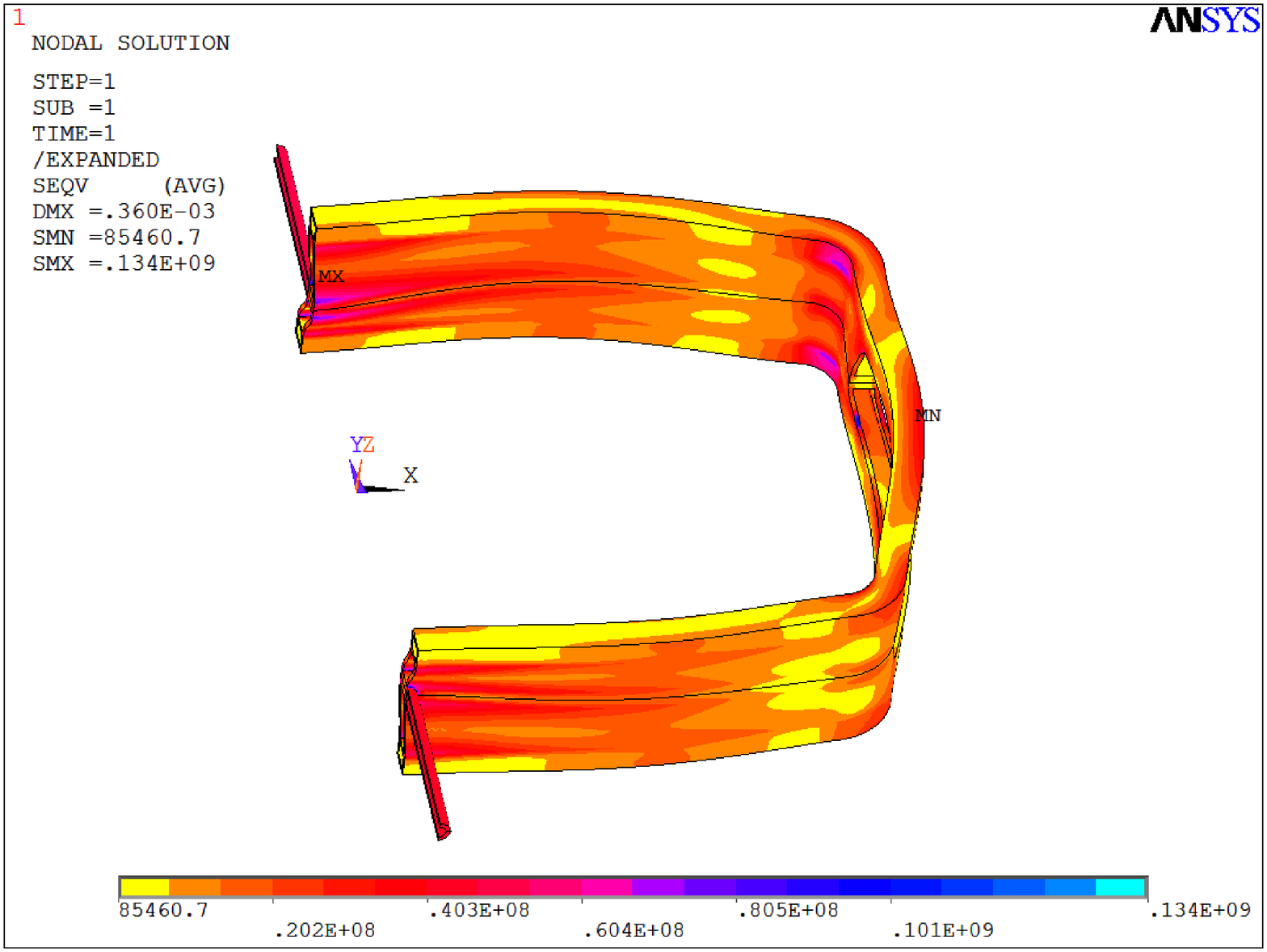}
\figcaption{\label{fig1}The stress distribution of coil case with supporters }
\end{center}
\begin{center}
\includegraphics[width=8cm]{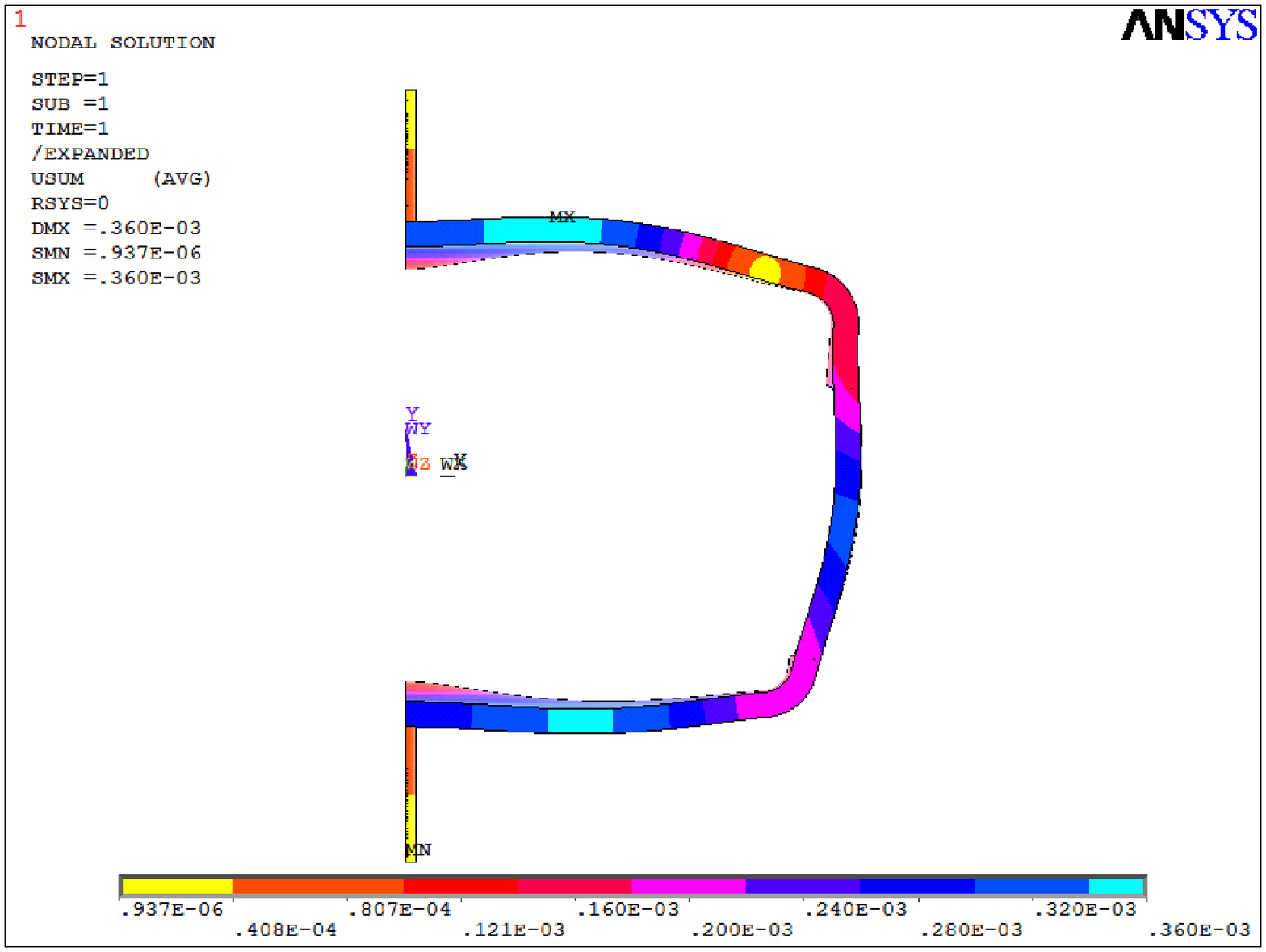}
\figcaption{\label{fig2}The displacement distribution of coil case with supporters }
\end{center}

With the constraint of supporters in the center of long edges, the maximum effective stress in the corner of coil is $18$ MPa, and the maximum stress of coil case is $134$ MPa. The overall deformation of coil and coil case is reduced to $0.5$ mm which could be accepted. The stress and deformation distribution of coil case with supporters constraint are shown in Fig.~\ref{fig1} and Fig.~\ref{fig2}.

\section{The thermal structural analysis in the cooling down process}

In the process of cooling down, the thermal contraction of different parts of coil system will cause large deformation and thermal stress. The winding coil is set in the coil case after vacuum pressure impregnation (VPI), after cooling down of superconducting coil, the gap between coil and coil case is about $0.2$ mm which could be neglected, and the thermal deformation of coil case along the long edge is about 6 mm. Therefore, the adjustable supporters should be taken into account in the structural design.

The supporters connect the coil case ($4.2$ K), thermal shield ($80$ K) and cryostat (room temperature) together. The thermal conduction is one of the major factors to cause the heat load and temperature gradient distribution. The deformation of supporter is more complex in condition of cooling down. In the stress analysis of supporter, it is supposed that the supporters end which connect the coil case is constrained and it is no displacement happen, the stress distribution on the most part of supporter is less than $100$ MPa, and the displacement in the thermal contraction along the axial direction is $2.6$ mm. The Fig.~\ref{fig3} and Fig.~\ref{fig4} show the stress and deformation distribution of supporter when cooled down. If the supporters ends are strictly fixed on both the $4.2$ K and $300$ K sides, the thermal stress will be huge ($900$ MPa) and beyond the ultimate strength of material, and so, the supporters should be adjustable to avoid the thermal stress during cooling down.
\begin{center}
\includegraphics[width=8cm]{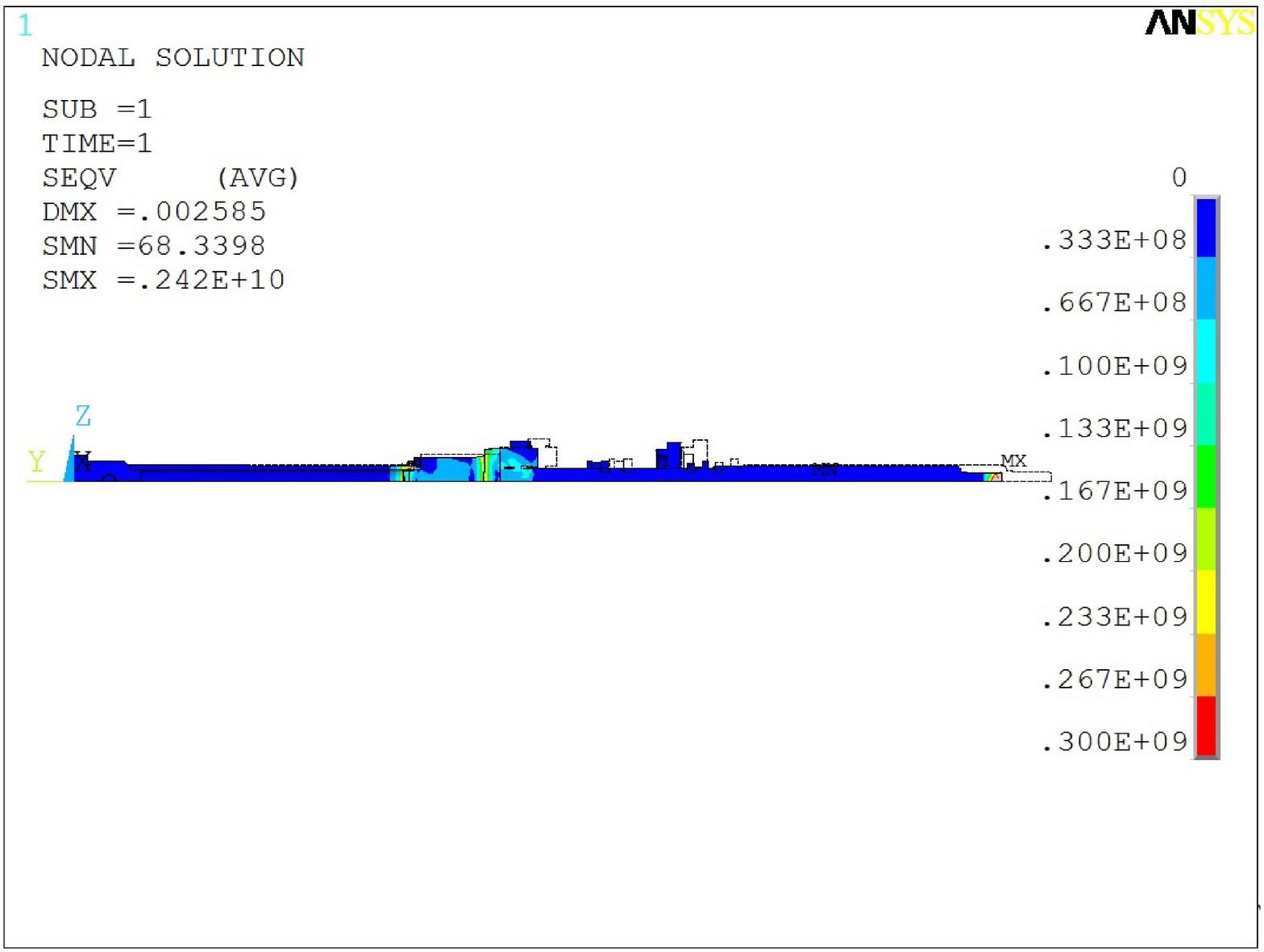}
\figcaption{\label{fig3}The supporter's stress distribution of cooling down (Constraint at low temperature end) }
\end{center}
\begin{center}
\includegraphics[width=8cm]{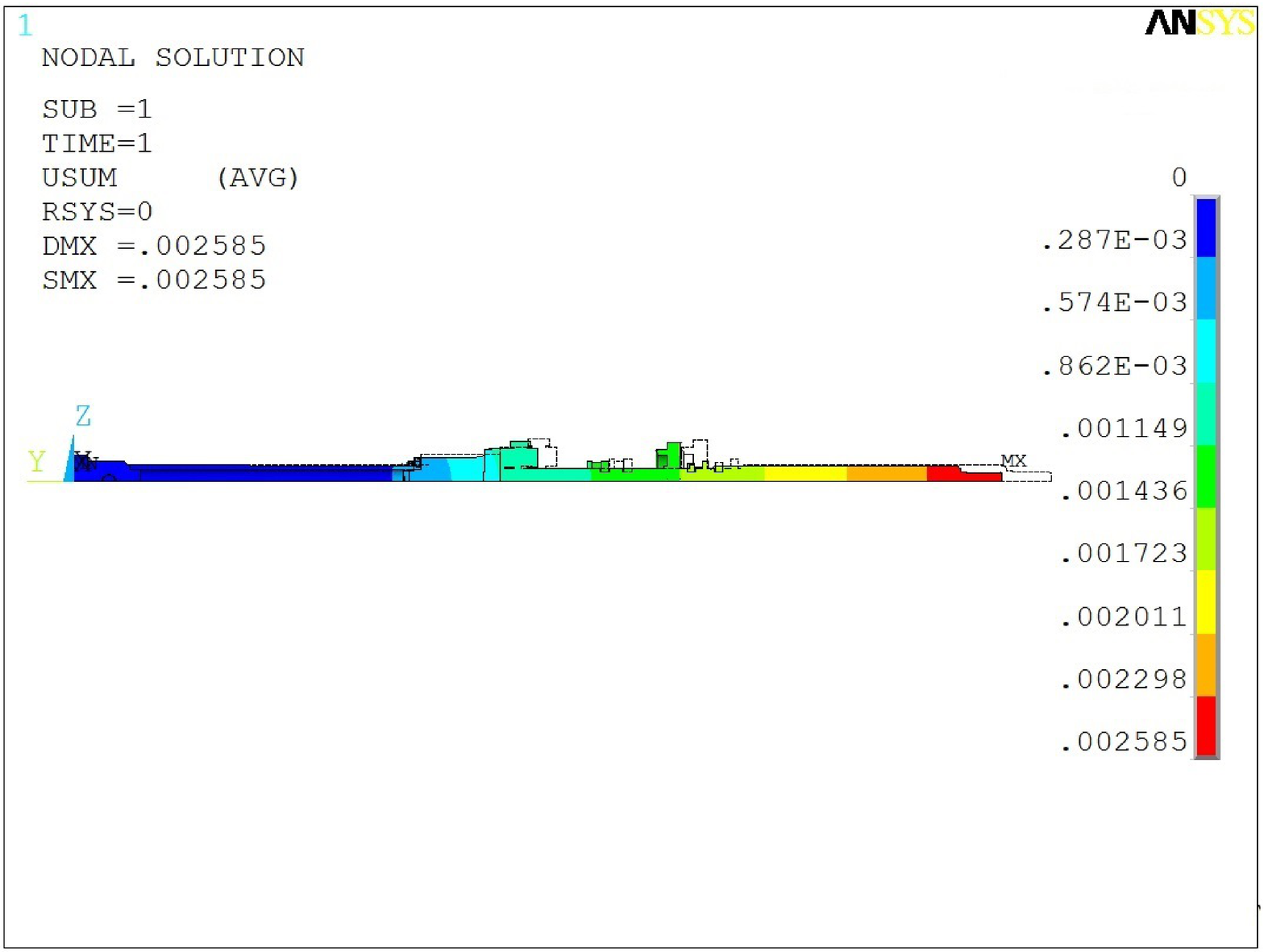}
\figcaption{\label{fig4}The supporter's deformation distribution of cooling down (Constraint at low temperature end) }
\end{center}

\section{Deformation of cryostat during pumping}

The cryostat is made of $316L$ stainless steel with thickness $10$ mm. Two situations are considered in condition of atmosphere pressure, the first one is without supporter and any constraints. The maximum stress appears in the center of long edge of cryostat, the maximum stress is $100$ MPa and the deformation of cryostat is about $1.4$ mm for each side. The stress distribution of cryostat during pumping is shown in Fig.~\ref{fig5}. In the practical structure of coil system, the cryostat is connected with coil case by supporter. For the constraint of supporter, the deformation of cryostat in place of connection will be reduced to $0.5$ mm. The deformation distribution of cryostat is shown in Fig.~\ref{fig6}. Once pumped, the cryostat deformation will not happen again and again, the cryostat with supporter could satisfy the structure design.
\begin{center}
\includegraphics[width=8cm]{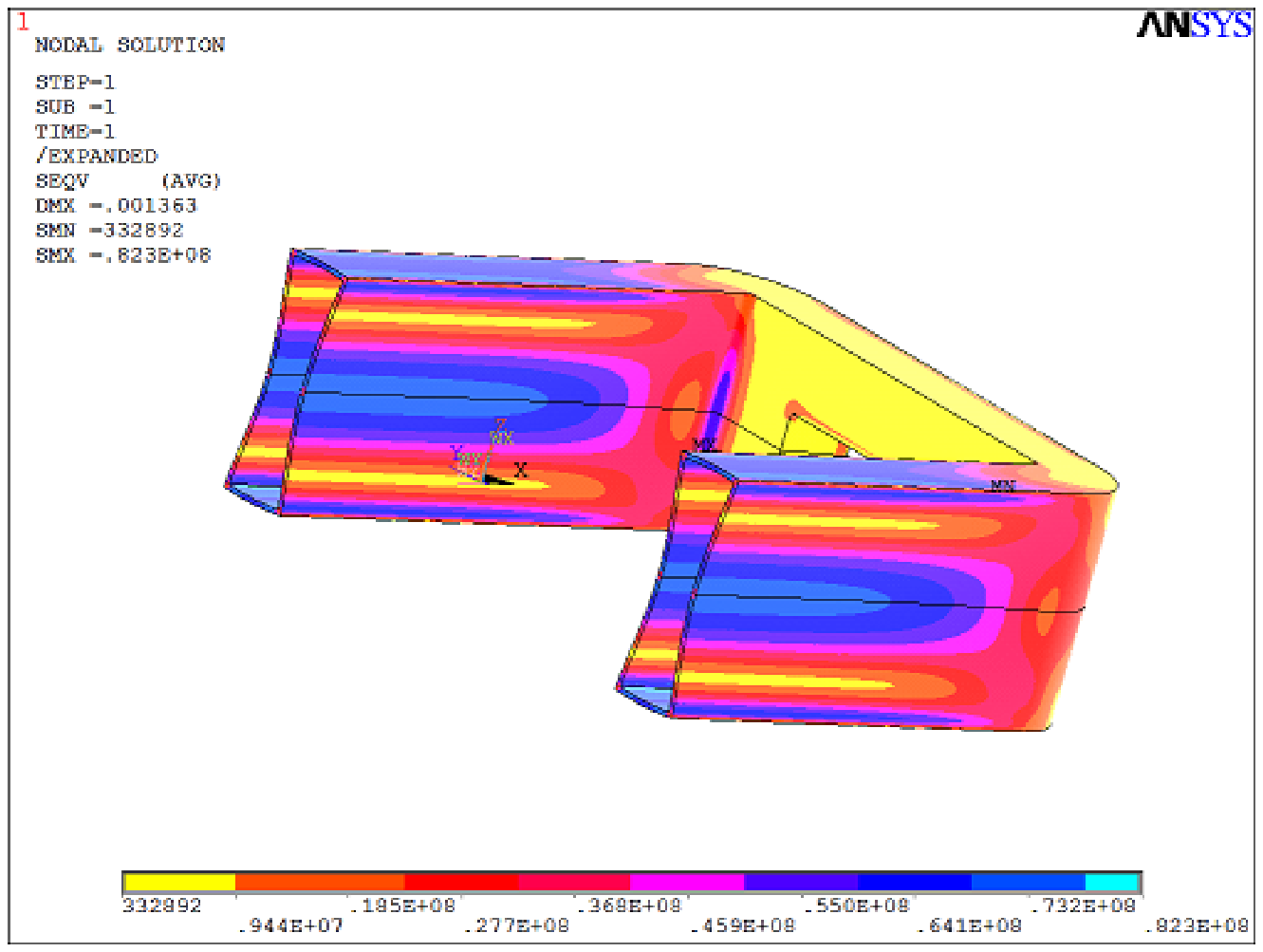}
\figcaption{\label{fig5}The stress distribution of coil case with supports }
\end{center}
\begin{center}
\includegraphics[width=8cm]{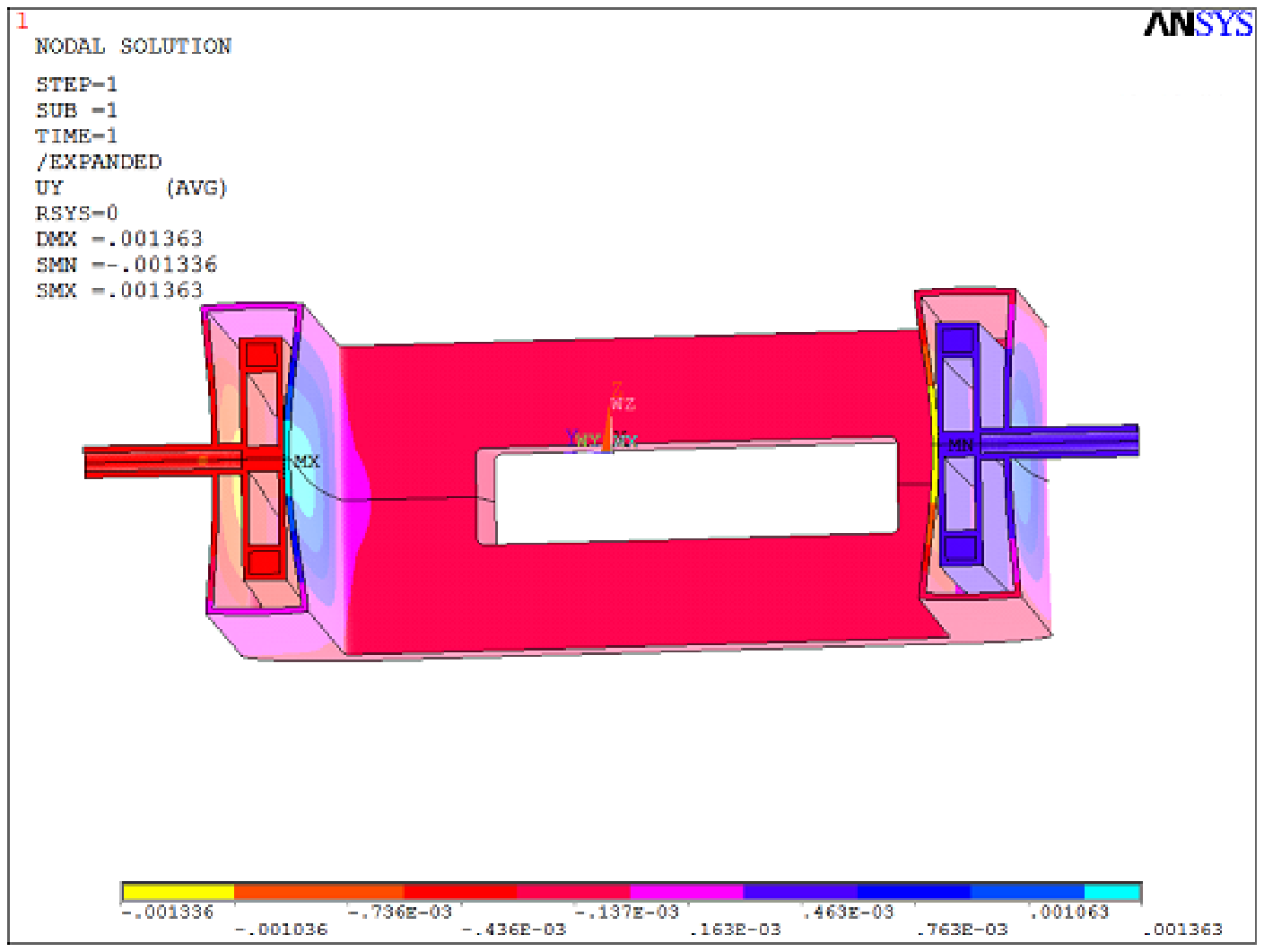}
\figcaption{\label{fig6}Deformation of cryostat with supporters during pumping }
\end{center}

\section{The structure design of supporter}

According to the mechanical analysis of coil case and cryostat in the different conditions, the supporter is very important for the whole structure design of coil system. The whole structure and the section structure of coil system are shown in Fig.~\ref{fig7} and Fig.~\ref{fig8}.

\begin{center}
\includegraphics[width=8cm]{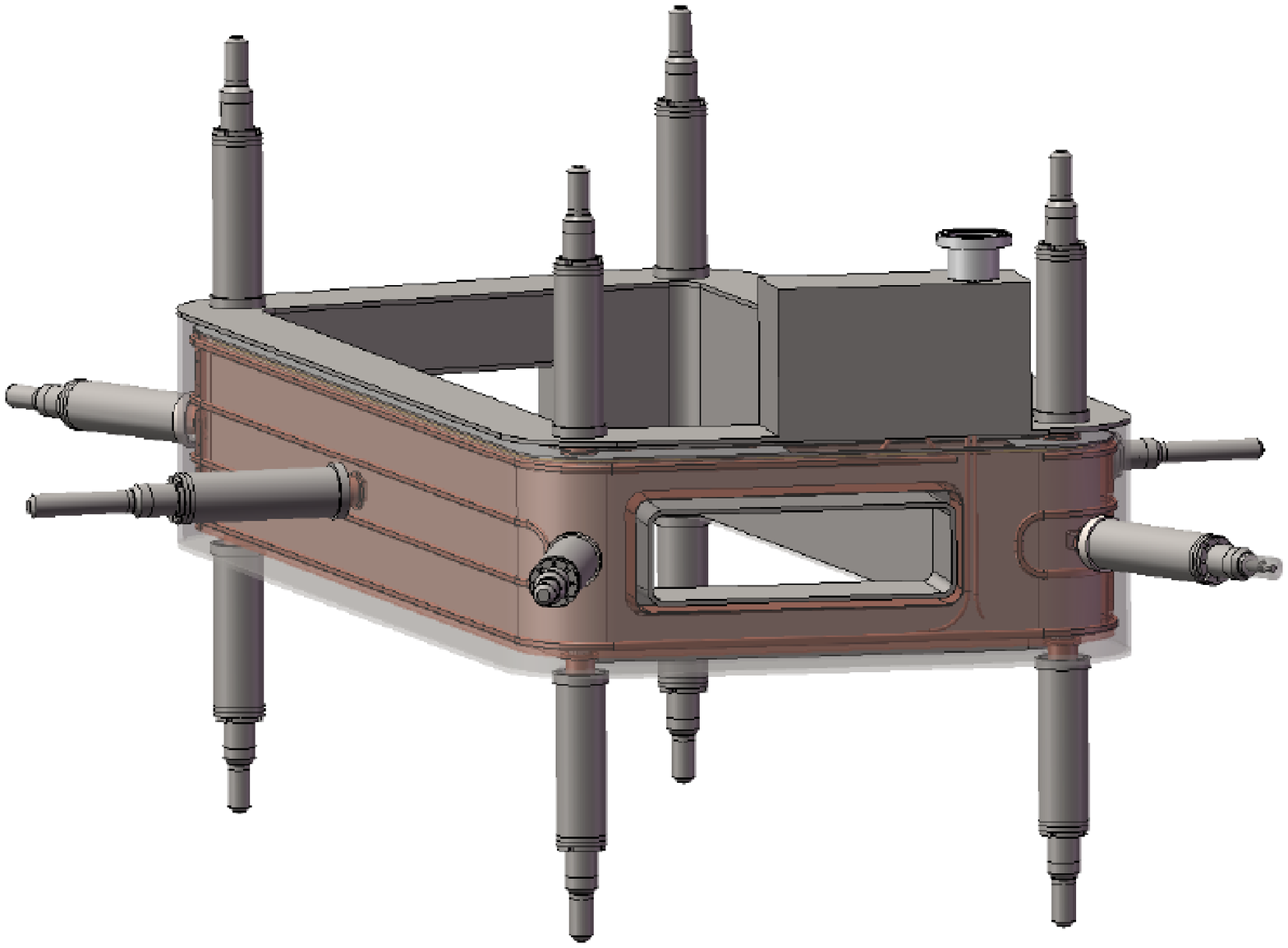}
\figcaption{\label{fig7}The structure of superconducting coil system }
\end{center}
\begin{center}
\includegraphics[width=8cm]{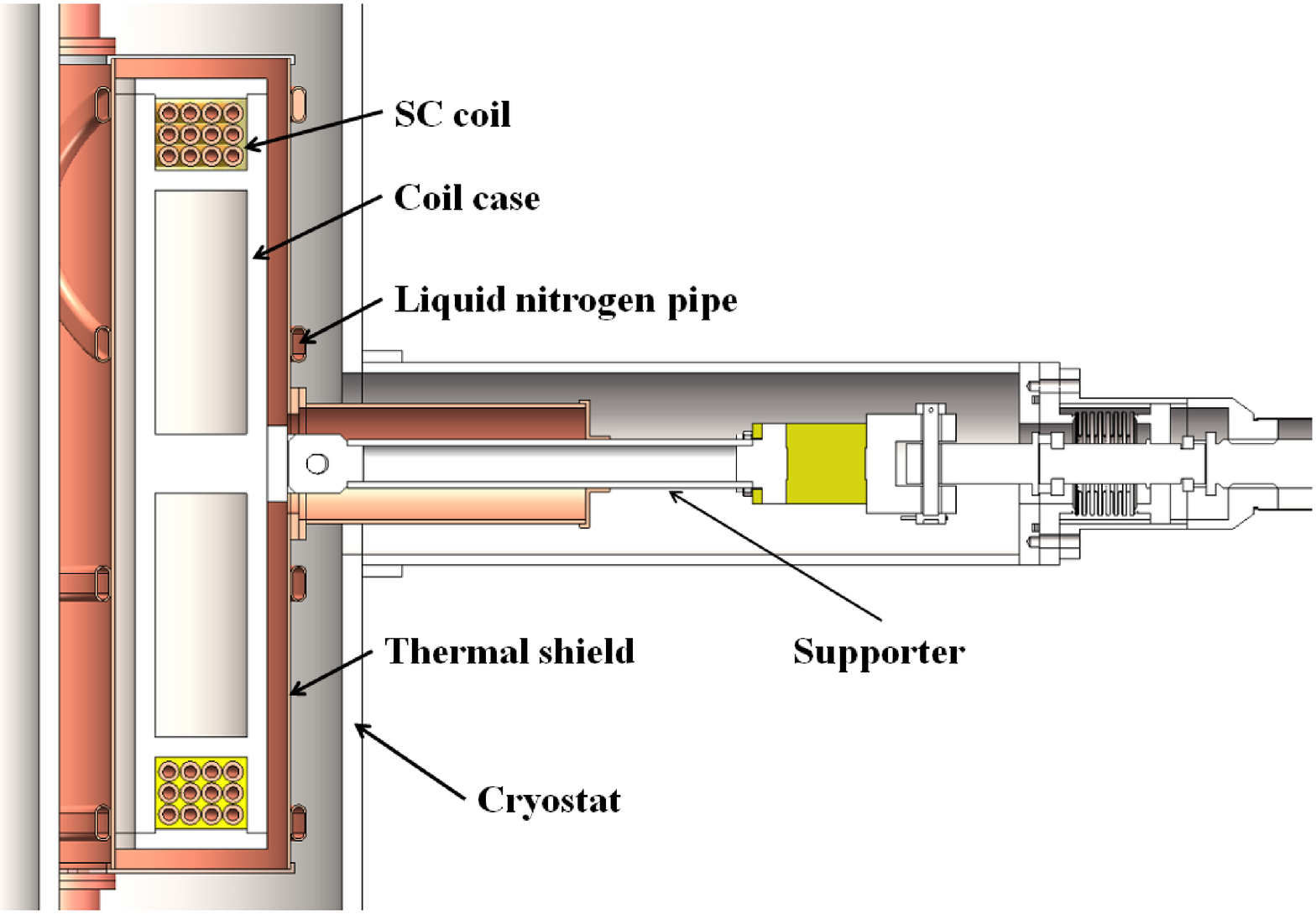}
\figcaption{\label{fig8}The section structure of superconducting coil system }
\end{center}

There are total of $14$ supporters used in the coil system, $12$ supporters are used for coil case fixation and position adjustment, and $2$ supporters in the middle of long edges of coil case are used to resist the electromagnetic force. For the reason of connection with coil case, thermal shield and vacuum seal, the material with good strength and low heat conductivity is required in the practical structure design of supporters. For $12$ supporters in the coil corner, G$10$ rod is used for its good heat insulation property. The reasonable position on the rod is chosen to connect the thermal shield to minimize the temperature transfer from $4$ K to $80$ K while maintaining reasonable heat load from room temperature.  At the supporters room temperature end, the flexible bellows is used for vacuum seal and position adjustment. For the supporter in the middle of long edge of coil case, the titanium alloy pipe is used for good strength and heat insulation than stainless steel. The warm end of supporter is connected with yoke for constraint to transfer the electromagnetic force to yoke and make sure the deformation of coil and coil case within reasonable range.

\section{Results}
 This paper analyzes the structure of superconducting coil system with finite element software ANSYS for the HIAF dipole prototype. According to the magneto-structural analysis and the thermal stress analysis of coil and coil case, and the mechanical stress analysis of cryostat during pumping, the optimal design of coil system is achieved. The reasonable structure of supporters could reduce the deformation of coil system effectively in operation. Recently, the dipole coil prototype has finished fabrication. Next, it will be tested to verify the reasonability of structure design.

\vspace{-1mm}
\centerline{\rule{80mm}{0.1pt}}

\end{multicols}

\clearpage

\end{document}